# Molecule/Electrode Interface Energetics in Molecular Junction: a "Transition Voltage Spectroscopy" Study

*Guillaume Ricœur, Stéphane Lenfant*, David Guérin, Dominique Vuillaume**

Institut d'Electronique Microélectronique et Nanotechnologie (IEMN), CNRS,

University of Lille, B.P. 60069, Avenue Poincaré, F-59652, Villeneuve d'Ascq, France

**ABSTRACT**: We assess the performances of the transition voltage spectroscopy (TVS) method to determine the energies of the molecular orbitals involved in the electronic transport though molecular junctions. A large number of various molecular junctions made with alkyl chains but with different chemical structure of the electrode/molecule interfaces are studied. In the case of molecular junctions with "clean, unoxidized" electrode/molecule interfaces, i.e. alkylthiols and alkenes directly grafted on Au and hydrogenated Si, respectively, we measure transition voltages in the range 0.9 – 1.4 V. We conclude that the TVS method allows estimating the onset of the tail of the LUMO density of states, at energy located 1.0 – 1.2 eV above the electrode Fermi energy. For "oxidized" interfaces (e.g. the same monolayer measured with Hg or eGaIn drops, or monolayers formed on a slightly oxidized silicon substrate), lower transition voltages (0.1 - 0.6 V) are systematically measured. These values are explained by the presence of oxide-related density of states at energies lower than the HOMO/LUMO of the molecules. As such, the TVS method is a useful technique to assess the quality of the molecule/electrode interfaces in molecular junctions.

## 1. INTRODUCTION

Over the past decade, numerous fabrication and measurement techniques have been developed to characterize the charge transport in metal/Self Assembled Monolayer (SAM)/metal junctions including scanning probes (STM, C-AFM), mechanically-controlled break junctions (e-beam lithographed and STM-based), nanopore-based junctions, metallic wire cross-bars, top electrode deposited by transfer printing, or using “liquid” contact (e.g. Hg drop) - (for a review see reference[1]). This abundant literature on the subject shows an important dispersion of the measured conductance for the same type of molecules.[2] This dispersion is related to the fact that the conductance of the molecular junctions is strongly sensitive to the chemical nature and structural details of the molecule/electrode interfaces and to the molecular organization in the monolayer. These last few years, the “Transition Voltage Spectroscopy” technique (or TVS)[3;4;5;6;7] has been use more frequently to estimate the energy barrier height at the electrode/molecule interface (i.e. the energy offset between the Fermi energy of the metal electrode and one of the molecular orbitals of the molecule). In this method, the energy barrier height is directly estimated from current-voltage (I-V) measurements, by plotting the I-V data in the form of a Fowler-Nordheim plot ($\ln(I/V^2)$ vs. $1/V$). In the classical interpretation of electron transport through a tunneling barrier,[4;8] the voltage at which a minimum is observed in this plot represents the transition voltage $V_T$ between the direct and Fowler-Nordheim tunneling regime. Applied to molecular junctions, it is shown that $V_T$ can give an estimation of the energy position of the molecular orbital (relative to the Fermi energy of the electrodes) involved in the transport mechanism, via a simple relation-ship $\Phi = \alpha V_T$, where $\alpha$ depends on several device parameters (symmetry of the junction in particular) ($0.8 < \alpha < 2$).[9;10] Albeit, the fact that the exact value of $\alpha$ and the physical origin of $V_T$ are still under debate,[9;10;11;12] TVS has become an increasingly popular tool in molecular electronics.

*Beebe et al.*[3;4] compare the $V_T$ measured on SAM of alkylthiol molecules by two techniques: conducting-AFM (alkyl chains with 6 to 12 carbon atoms) and cross wires (12 to 18 carbon atoms). They observe a constant $V_T$ with the length of the molecules: $V_T = 1.22 \pm 0.05$ V. Another work[6] on alkyldithiol molecules characterized by electromigrated nanogap junction at 4.2 K reports also a constant $V_T = 1.9 \pm 0.1$ V with length of molecules for 8 to 12 carbon atoms. Using STM-based break junctions with alkyldithiols, *Guo et al.*[13] report $V_T$ in the range 1 - 1.5 eV. Recently, *Clement et al.*, also report $V_T$ around 1.2 – 1.8 V for junctions made of a small amount (~ 80) of alkylthiols (12 carbon atoms) grafted on a tiny monocrystalline Au nanodot (< 8 nm in diameter) contacted by a C-AFM tip.[14]

Albeit the TVS approach is mainly applied on molecular junctions with thiolated molecules on gold substrate, one group[15;16] reports very small $V_T$ (about 0.2 V) for alkyl molecules on silicon substrates. Electron transport through silicon/molecule interface states is suggested to explain this low value.

Here, to assess the performances and potential of the TVS method, we extensively use this method on a large number of molecular junctions with different type of molecule/electrode interfaces, different test-beds and different lateral sizes (from few tens of nm to mm). As top contact of the molecular junctions, we use various techniques: conducting-AFM,[17] aluminum evaporation through a shadow mask,[18] “liquid” contact by eutectic GaIn[19] or mercury[20] drop and micropore-based junctions (Figures 1a to 1e). We apply these contacting techniques to three families of SAM: alkylthiol ($CH_3$-$(CH_2)_{n-1}SH$ with n=4 to 18) on gold surfaces, alkyltrichlorosilane ($CH_3$-$(CH_2)_{n-1}SiCl_3$) with n=8 to 12 on silicon with native oxide and alkene ($CH_2$=CH-$(CH_2)_{n-3}$-$CH_3$ with n=6 to 10) grafted on hydrogenated silicon (Figure 1f). As a result, 10 different types of molecular junctions (with different chemical nature of the electrode/molecule interface, lateral size, and chain length) are characterized. The chosen molecules, based on alkyl chains, constitute an interesting case study for TVS, because (i) their HOMO (Highest Occupied Molecular Orbital) - LUMO (Lowest Unoccupied Molecular Orbital) gap remains constant with the number of methylene units in the chain (as soon as larger than 4)[21;22], (ii) the electronic transport in the junction is a non resonant tunneling[21]. In many previous works, only one bias polarity is used to determine $V_T$. In this study, we determine systematically $V_{T+}$ at positive bias and $V_{T-}$ at negative bias, since these values can be used to determine the energy position of the molecular orbital involved in the electrical transport[11] and the degree of symmetry/asymmetry of the junction.[23] We also discuss and compare the TVS results with independent measurements of the energetics of the molecular junctions (by UPS and IPES) when such data are available in literature for the same type of molecular junctions.

## 2. EXPERIMENTAL METHODS

We fabricated several molecular junctions belonging to three different groups. The first group referred to as "oxide free" junctions corresponds to SAMs of alkylthiol on gold and alkene on hydrogenated silicon substrates, contacted with a gold C-AFM tip: $Au_{sub}$-S-$C_nH_{2n+1}$//$Au_{tip}$ (n=4,6,8,12,14,16,18), and Si-$C_nH_{2n+1}$//$Au_{tip}$ (n=6,8,10), where "-" denotes an interface with a chemical bond, and "//" a non-covalent, or mechanical interface. The second group referred to as "oxidized top contact" corresponds to the same SAMs contacted with eutectic GaIn (eGaIn), Hg drop or evaporated Al contacts. In this group, we measured 4 different types of junctions: $Au_{sub}$-S-$C_nH_{2n+1}$//$Ga_2O_3$/eGaIn, Si-$C_nH_{2n+1}$//$Ga_2O_3$/eGaIn, Si-$C_nH_{2n+1}$//HgO/Hg and Si-$C_nH_{2n+1}$/$Al_2O_3$/Al, where "/" denotes an intimate contact at the interface, e.g. the interface between a metal and its oxide, or between a SAM and an evaporated metal. Both eGaIn and Hg drops have a very thin superficial oxide, likely $Ga_2O_3$ (~7 Å) and HgO (~16 Å) respectively[24;25]. In the case of evaporated Al electrodes, it is likely that oxygen (residual oxygen in the vacuum chamber during evaporation, and/or when exposing the samples to air) can form some oxide (likely $Al_2O_3$) at the interface as observed by infrared spectroscopy on silane junctions[26]. The last group, referred to as "both oxidized interfaces" is based on alkyltrichlorosilane SAMs on slightly oxidized silicon substrate, and consists of three types of junctions: Si/$SiO_2$-Si-$C_nH_{2n+1}$//$Ga_2O_3$/eGaIn, Si/$SiO_2$-Si-$C_nH_{2n+1}$//HgO/Hg and Si/$SiO_2$-Si-$C_nH_{2n+1}$//$Al_2O_3$/Al (n = 8, 12).

### 2.1 Chemicals and SAM fabrication

For alkylthiol ($CH_3$-$(CH_2)_{n-1}$SH with n = 4, 6, 8, 12, 14, 16, 18 from Aldrich) SAMs, we used a Si wafer <100> n-type single side polished as mechanical substrate. It was cleaned with solvents (acetone and isopropanol VLSI grade from Carlo Erba) under ultrasonic bath and dried under nitrogen flow before deposition by sputtering of 10 nm-thick of titanium and 100 nm-thick of gold on substrate heated at 350°C. Then, the gold-coated substrate was directly immerged in a freshly prepared 3 x $10^{-3}$ M solution of the alkylthiol in ethanol for a minimum of 48h. The samples were rinsed in an ultrasonic bath with solvents (acetone followed by isopropanol) before use.

For alkene monolayers ($CH_2$=CH-$(CH_2)_{n-3}$-$CH_3$ with n = 6, 8, 10 from Aldrich purity >98%), we used a highly-doped Si substrate as bottom electrode. The Si wafer <111> p-type single side polished (resistivity < 0.01 Ω.cm) was cleaned by rinsing with solvents and blown dry under a stream of argon. Subsequently, the wafer pieces were immersed in piranha solution (98% $H_2SO_4$: 30% $H_2O_2$, 2:1, v/v) at 90°C for at least 30 min, rinsed with copious amounts of deionized MilliQ wafer and etched in deoxygenated 40% $NH_4F$ solution for ~15 min. This treatment (immersion in piranha solution, rinsing, and etching in 40% $NH_4F$ solution) was repeated once (*caution: piranha solution is very exothermic and reactive with organics; it should be handled with extreme care*). Alkyl monolayers were formed via thermal hydrosilylation of alkenes.[27;28;29] The freshly etched piece of Si wafer was immersed in neat deoxygenated alkene under argon and heated at 200°C for 4h (for n = 18, 14) (or respectively 150, 100 and 70°C for n = 10, 8, 6 in order to reduce the evaporation short-chain alkenes). From XPS measurements, we checked that no silicon oxide peak is detected just after the SAM formation. Samples were immediately stored under glove box filled with nitrogen and electrically characterized less than few days after fabrication.

The alkyltrichlorosilane molecules ($CH_3$-$(CH_2)_{n-1}SiCl_3$ with n = 8, 12 from Aldrich) were chemisorbed on naturally oxidized silicon substrates (1 to 1.5 nm thick $SiO_2$ as measured by ellipsometry) from a dilute solution ($10^{-3}$ M) in an organic solvent (70/30 % v/v of hexane and carbon tetrachloride) using the method developed by Maoz and Sagiv[30] and later improved by other groups[31]. The highly-doped (degenerated) n-type <100> silicon substrates (resistivity of ~ $10^{-3}$ Ω.cm) were carefully cleaned by a piranha solution ($H_2SO_4$:$H_20_2$ 2/1 v/v) followed a dry UV-ozone treatment. The cleaned substrates were dipped into the freshly prepared solution during 90-120 minutes. The formation of the SAM is highly sensitive to traces of water. For a better control and a good reproducibility, we worked in a glove box, maintained under a dry and clean nitrogen circulation (relative humidity less than 0.1 ppm). After the SAM formation, the samples were rinsed in an ultrasonic bath during 5 min with solvents (chloroform followed by isopropanol, VLSI grade from Carlo Erba).

For all these monolayers, we measured the thickness (ellipsometry) and the water contact angle. These values (see Supporting Information) were compared to literature data such as to check that we have fabricated SAMs with a good structural quality (densely compact monolayer).

**2.2 Aluminum electrode and micropore junction fabrication**

For the electrical measurements, we formed molecular junctions by evaporating metal (aluminum) through a shadow mask (electrode area $10^{-4}$ $cm^2$). To avoid contaminating of the surface during metallization, an ultra-high vacuum (UHV, at $10^{-8}$ torr) electron-beam evaporation system was used. In order to avoid, as much as possible, damage of the SAM during deposition a low evaporation rate at 3 Å/s was used to form electrodes with a thickness of 40 nm. We get a fabrication yield of 95 % (i.e. percentage of non short-circuited junctions).

The micropores were fabricated in thermally oxidized ($SiO_2$ 100 nm thick) <100> silicon with a low resistivity (~$10^{-3}$Ω.cm, n-type). Standard e-beam lithography with 200 nm-thick PMMA resist (4% 950K) was used, micropores (diameter of 10 and 100 µm, spaced by 350 µm) were etched in the silicon dioxide by reactive ion etching (RIE $CHF_3$ 20 sccm/ $CF_4$ 20 sccm, 50 mT, 80 W during 8 min). After removing the resist (with acetone) and wafer cleaning by UV-ozone the alkylsilane SAMs were prepared by the same procedure as detailed before. Then a 100 nm-thick aluminum layer was deposit on the wafer at 3 Å/s in an UHV electron-beam evaporation system as above. Finally top electrodes were patterned (i.e. isolated from each other) by scratching the aluminum layer around the measured micropore with a tip controlled by a micromanipulator.

### 2.3 C-AFM measurement

Current-voltage measurements were performed by conducting atomic force microscopy (C-AFM) under a nitrogen flow (Dimension 3100, Veeco), using gold-coated tip (from MikroMasch reference CSC17, 0.15 N/m spring constant). To form the molecular junction, the conducting tip was localized at a stationary contact point on the SAM surface at controlled loading force (around 20 nN, otherwise specified). Current–voltage (I-V) characteristics were acquired and treated with the Nanoscope 5.30R2 software from Veeco Instrument Inc. The I-V characteristics were obtained directly by varying voltage from -1.5V/-2V to +1.5V/+2V (depending on breakdown field supported by the various samples) and reversibly, at different places on the sample, without averaging between successive measurements. Moreover, due to the small contact area (about 10 $nm^2$ at 20 nN), the measured current is very low (between 100 nA and few pA) and imposes a limit for the SAM thickness (no more than 12 carbon atoms in general). The voltage was applied on the substrate, the tip being ground (i.e. at the input of current amplifier). However, note that for the sake of comparison with eGaIn, Hg drop, Al electrodes, all I-V curves are presented in the figures of this paper with the voltage applied on the top electrode (i.e voltage scale has been inverted for C-AFM measurements).

### 2.4 eGaIn and Hg drops contacts

To form molecular junctions, we also used eutectic Gallium Indium drop contact (eGaIn 99.99%, Ga:In; 75.5:24.5 wt% from Alfa Aesar). We used a method close to the one developed by *Chiechi et al.*[32]. We formed a drop of eGaIn at the extremity of a needle fixed on a micromanipulator. By displacing the needle, we brought the drop into contact with a sacrificial surface, and we retracted the needle slowly. By this technique, we formed a conical tip of eGaIn with a diameter of around 200 µm (corresponding to contact area of ~ $10^{-3}$ $cm^2$). This conical tip was then put into contact with the SAM (under control with a digital video camera). We used the eGaIn tip at different places on the samples and we regularly formed a new tip for a same sample to avoid pollution of the ip. The eutectic GaIn drop allows a good electric contact with the SAM since eGaIn accommodates well the surface roughness, and can be used on gold without forming amalgam.[32]

We also used a hanging mercury drop, to contact the SAM. Calibrated mercury drops (99.9999%, purchased from Fluka) were generated by a controlled growth mercury electrode system (CGME model from BASi). The mechanical contact between the sample and the hanging mercury drop was formed my moving up a precison lab-lift (supporting the sample) under the control of a digital video camera. The electrical contact area estimated by capacitance measurement is around $10^{-4}$ $cm^2$. This technique is only used for SAMs on Si substrates.

**2.5 I-V and TVS measurement protocol**

For eGaIn, Hg drop, micropore and aluminum pads, the current-voltage measurements were done with an Agilent 4156C parameter analyzer, the voltage was always applied on the top electrode and the current measured at the grounded substrate. As for C-AFM, voltage was varied from -1.5 V to +1.5 V and reversibly from +1.5 V to -1.5 V, at different places on the sample, without any averaging between successive measurements. Five to more than 30 I-V were measured for each junctions made with the different molecules, top contacts and chain length. This number is sample dependent and depends on to the removal of I-V curves characterized by the presence of short-circuits (very large current density > 1 $A/cm^2$) or poor electrical contact (i.e. with large noise or sudden and random step-like current fluctuations). Voltages at which a minimum in the $\ln(I/V^2)$ vs. 1/V were determined for each measurement and averaged to give the transition voltages for positive bias ($V_{T+}$) and for negative bias ($V_{T-}$) as well as the standard deviation (reported as the error bars in the figures) for each kind of junctions. For samples with a number of I-V measurements smaller than 10, we simply used the average value, for samples with a larger number of data (C-AFM), we plotted $V_T$ histograms, which were fitted by a Gaussian distribution.

## 3. RESULTS ANS DISCUSSIONS

### 3.1. Oxide-free junctions: Alkylthiol on Au and Alkene on Si measured by C-AFM

We measured current-voltage (I-V) curves for oxide free junctions with the following structure: freshly evaporated gold substrate/SAM/gold covered AFM-tip (see Experimental Methods). A typical set of 19 I-V curves for the $Au_{sub}$-S-$C_nH_{2n+1}$//$Au_{tip}$ (n=4) is shown in Fig. 2-a, and the corresponding $\ln(I/V^2)$ vs. 1/V in Fig. 2-b. Fig. 2-c shows the $V_T$ histograms obtained from Fig. 2-b, from which we get the $V_{T+}$ and $V_{T-}$ (and standard deviation) values for this junction (see experimental methods). The $V_{T+}$ and $V_{T-}$ values for junctions made of alkylthiols with different chain length ($Au_{sub}$-S-$C_nH_{2n+1}$//$Au_{tip}$) are presented in figure 3. The values for $V_{T+}$ and $V_{T-}$ are independent of the alkylthiol chain length in the SAM (n=4 to 12). Note that for n>12 (for the alkylthiol), no current can be measured due to the small contact area of the C-AFM tip. These $V_{T+}$ values are in good agreement with Beebe *et al.*[4], who reported values between 1.20 ± 0.08 V and 1.28 ± 0.12 V, measured on the same kind of junctions (n from 6 to 12) with the C-AFM technique.

Albeit TVS measurements for both bias polarity are sparsely reported, $V_T$ should strongly depend on the bias polarity due to the asymmetry of the junction, especially if the two electrode/organic interfaces are different.[3;4;33;34] These asymmetric junctions present unequal voltage drop at the two interfaces inducing a difference between the absolute values of $V_{T+}$ and $V_{T-}$. This feature was supported by *ab initio* DFT transport calculations simulating TVS curves for various molecular junctions[23].

Here, we systematically measured TVS for both bias polarities, in order to figure out if there is a general trend relating the structure (symmetric vs. asymmetric) of the junctions and the ratio $V_{T+}/V_{T-}$. The $V_{T+}$ and $V_{T-}$ values are shown in Fig. 3 and Table 1. Considering the error bars, the measured transition voltages are equal in absolute values for both polarities. For alkylthiol junctions, the averaged values of the transition voltages measured for each chain lengths are $V_{T+ave}$= 1.32 ± 0.08 V and $V_{T-ave}$= -1.26 ± 0.08 V (table 1), thus a TVS asymmetry ratio:

$$\text{TVS-AR}=\max\left(\left|\frac{V_{T+}}{V_{T-}}\right|,\left|\frac{V_{T-}}{V_{T+}}\right|\right) \qquad (1)$$

is of about 1.06 ± 0.13. Other authors[35] measured TVS on molecular junctions with nonanedithiol and decanethiol SAMs. The $V_{T+}$ values are the same for both molecules, about 1.1 ± 0.07 V. Recently, values in the range 1.1 - 1.5 V were reported for 8 and 10 carbon atoms alkyldithiol single molecule using STM based break junction methods, irrespective to the bias polarity, i.e. with a TVS-AR between 1 and 1.2[13]. Albeit the test-bed structures are not the same, and thus both the position of the molecular orbitals and the electronic coupling with electrodes can change from experiment to experiment, all these experiments show that the transition voltages are in the same range, with a rather symmetrical behavior for both bias polarities.

The loading force applied on the C-AFM tip is an important parameter when measuring the transport properties of SAM junctions. In general, the loading force is chosen arbitrary between few nN and few tens of nN.[3;4] However, the loading force affects radically the current[17;36;37] which increases with the loading force (e.g. the current measured on C8 alkylthiol is multiplied by more than $10^2$ when the force increases from 2 to 10 nN)[36]. We checked the effect of the loading force from 10nN to 40 nN on $V_T$ for alkylthiol (n=12) junctions (Figure 4). In this force range, we do not observe any influence. The $V_T$ values are almost constant for both bias polarities, with averaged values of $V_{T+ave}$= 1.23 ± 0.03 V and $V_{T-ave}$= -1.25 ± 0.06 V. *Wang et al.* [34] report a decrease of the transition voltage in the range 50 - 100 nN, which is attributed to a disorder-induced enhancement of the electronic transport (i.e. enhancement of the intermolecular chain-to-chain tunneling when increasing the molecule tilt by the applied force). Such a disorder related effect is not important here with loading forces smaller than 40 nN.

Figure 5 and table 1 show the TVS results for the Si-$C_nH_{2n+1}$//$Au_{tip}$ junctions. Again, we measured $V_T$ values independent of the chain length (above n=10, the current is too low to be measured by C-AFM). The average value is $V_{T-ave}$ = -0.96 ± 0.15 V. Note that we have not observed a transition voltage for the positive bias (except for n=8). This feature may be due to the fact that the measured current was too low for the positive bias, because these Si-$C_nH_{2n+1}$//$Au_{tip}$ junctions are more asymmetric than the alkylthiol one. This is confirmed by the larger (in absolute value) $V_{T+}$ of 1.4 V measured for the n=8 junction (TVS-AR ≈ 1.4).

### 3.2. Junctions with oxidized top electrode: Alkylthiol with eGaIn top contact, and Alkene with Hg, eGaIn or Al top electrode

We also characterized the same SAMs using eGaIn drop contact to compare the effect of the top electrode on $V_T$ values. The eGaIn drop presents on the surface a thin layer (*ca.* 0.7 nm) of gallium oxide (mainly $Ga_2O_3$) that contains many defects.[24] It is usually assumed that these defects may dope the oxide layer making it highly conducting. For this raison, the influence of this layer on the electrical properties of the molecular junctions is usually neglected.[24;38] For instance, *Reus et al.*[38] demonstrated that rectifying behavior in molecular junctions (ferocene and naphthoquinone derivatives) are due to the electronic properties of the SAM and not to the presence of $Ga_2O_3$ layer.[38] However, TVS data were not reported for these systems. Alkene monolayers on silicon were also characterized using a mercury drop and an aluminum electrode deposited by evaporation through a shadow mask. These materials present also a thin surface layer of oxide that forms when the metal is exposed to air, respectively HgO and $Al_2O_3$ (in this latter case, it is likely that $Al_2O_3$ is not only formed at the outer surface of Al, but also at the Al/monolayer interface, since O may diffuse through the organic SAM). These later two approaches cannot be applied for SAM on gold because mercury forms spontaneously amalgam with gold through the single monolayer (capping the Hg/HgO drop with a second SAM may be used to delay the amalgam formation,[39] but was not used here), and aluminum evaporated on SAM on Au causes 100% short circuited devices.

From the current-voltage characteristics measured with the eutectic GaIn drop on a series of alkylthiol $Au_{sub}$-S-$C_nH_{2n+1}$//$Ga_2O_3$/eGaIn junctions (4 to 18 carbon atoms) two $V_T$ values are deduced, one for each polarities, which are independent of the length of molecules (Fig. 3, table 1) as already observed above by C-AFM and by other groups.[3;6] The average value determined for all the $Au_{sub}$-S-$C_nH_{2n+1}$//$Ga_2O_3$/eGaIn junctions at positive and negative bias are respectively $V_{T+ave} = 0.44 \pm 0.12$ V and $V_{T-ave} = -0.52 \pm 0.06$ V. Considering uncertainties, $V_T$ values are equal (in absolute values) for both polarities as observed by C-AFM.

In figure 3, the comparison between eGaIn and C-AFM measurements shows the influence of the top electrode on the $V_T$ values. By changing the top electrode from gold (C-AFM) to oxidized GaIn (eGaIn/$Ga_2O_3$) the $V_T$ value decreases by *ca.* 0.8 - 0.9 V. The origin of this decrease can be due to two factors: (i) the reduction of the work function of *ca.* 1eV from gold to eGaIn,[19;24] (ii) or the presence of oxides states related to the $Ga_2O_3$ layer in the junction. In the first case, a reduction (in absolute value) of both $V_{T+}$ and $V_{T-}$ with a decrease of the work function implies that the tunneling transport is LUMO-mediated (see discussion section) with a nearly ideal symmetric junction (i.e. $\gamma = 0$, where $\gamma$ is a voltage division parameter describing the degree of symmetry or asymmetry of the molecular orbitals in the junction, $-0.5 \leq \gamma \leq 0.5$).[10;11] Albeit there is no data in literature for $\gamma$ in the case of molecular junction with eGaIn electrodes, DFT calculations predicted an almost symmetric coupling ($\gamma$ = 0.14 - 0.16) for the $Au_{sub}$-S-$C_nH_{2n+1}$//$Au_{tip}$ junctions.[10] In the second case, we assume that oxide states are introduced at energy smaller than the LUMO/HOMO gap, and that these states are now probed by the TVS method at lower bias than for the molecular orbitals. This point will be further discussed below (see discussion section).

In the case of alkene molecules directly grafted on silicon without native oxide and electrically contacted by top oxidized electrodes (mercury drop contact, eGaIn drop or aluminum patterns), I-V curves plotted in TVS form show clear and reproducible minima. The $V_T$ values for the three junctions, Si-$C_nH_{2n+1}$//$Ga_2O_3$/eGaIn, Si-$C_nH_{2n+1}$//HgO/Hg and Si-$C_nH_{2n+1}$/$Al_2O_3$/Al, are reported Fig. 5 and in table 1 for comparison with the Si-$C_nH_{2n+1}$//$Au_{tip}$ C-AFM junction. Again, the $V_T$ values are almost independent of the molecule length (if we except a slight increase for n= 14 and 18 in the case of the eGaIn contacts, but with a large data dispersion). Also, the $V_T$ values are almost symmetric for both polarities. We note that these values for the different junctions with the different top electrodes are quite similar. The fact that $V_T$ values are almost the same whatever the nature of the oxidized electrodes (averaged values of $V_{T+ave} = 0.27 \pm 0.14$ V and $V_{T-ave} = -0.37 \pm 0.18$ V for positive and negative bias, horizontal lines in Fig. 5) supports the hypothesis that TVS is measuring some oxide states in the junctions lying inside the LUMO/HOMO gap.

### 3.3. Junctions with both oxidized interface: alkyltrichlorosilane

Alkyltrichlorosilane SAMs grafted on slightly oxidized silicon were used to fabricate molecular junctions with two oxidized interfaces. Different electrodes were used to electrically characterized these junctions, all with oxidized metals (eGaIn, Hg and Al): $Si/SiO_2$-Si-$C_nH_{2n+1}$//$Ga_2O_3$/eGaIn, $Si/SiO_2$-Si-$C_nH_{2n+1}$//HgO/Hg and $Si/SiO_2$-Si-$C_nH_{2n+1}$//$Al_2O_3$/Al (n=8, 12). The presence of both a silicon oxide and the SAM does not allow using C-AFM, the current is too low. Again, $V_T$ values are independent of chain length, bias polarity, and size of the surface contact area (from $10^{-3}$ cm² to $10^{-6}$ cm² for respectively Hg, eGaIn drops and 10 µm and 100 µm diameter micropores) as summarized in Fig. 6 and Table 1. The $V_T$ values for all these alkyltrichlorosilane junctions are close to the previous values obtained on alkene SAMs with a top oxidized contact (see Fig. 5), with average values of $V_{T+ave}$ = 0.25 ± 0.10 V and $V_{T-ave}$ = -0.21 ± 0.08 V respectively for positive and negative bias (horizontal lines Figure 6). For $Si/SiO_2$-Si-$C_nH_{2n+1}$//$Al_2O_3$/Al junctions, LUMO and HOMO energy levels with respect to the Fermi energy of the electrodes were reported at ~ 4 - 4.5 eV by photoconductivity experiments.[21;40] Thus, we expect $V_T$ to be in the range of a few eV in that case. This major discrepancy between the $V_T$ and the barrier height estimated by another way shows the limitation of the TVS approach for these oxidized junctions. This is corroborated by $V_T$ measured on a reference junction (*i.e.* without SAM; aluminum, eGaIn or mercury directly deposited on silicon with its native oxide), as reported in figure 6 and in table 1 as n=0. The same $V_T$ values are measured with and without the SAM. Thus, it is likely that these low $V_T$ values correspond to oxide-related states located close to the electrode Fermi energy rather than related to the HOMO or LUMO of the molecules.

## 4. DISCUSSION

For the interpretation of these TVS results, we follow Araidi and Tsukada[12] who suggested, on the basis of non-equilibrium Green's function combined with both density-functional theory and tight-binding approximation, that the TVS transition voltage corresponds to the bias situation when the tail of the density of states (DoS) of the molecular orbitals (HOMO or LUMO) enters the energy window defined by the applied bias (see figure 7.a). We also assume that electron transport properties is controlled by the LUMO as recently demonstrated for both alkylthiol junctions on Au[41] and alkene on hydrogenated silicon.[42] Inverse photoemission spectroscopy (IPES) on C18/Au interface shows that the LUMO-edge is at 3.35 eV above the Au Fermi energy[41], far away the transition voltage of about 1.2 - 1.3 V. However, a clear DoS tail is also observed at lower energy, which is related to induced-density of interface states (IDIS) or metal-induced gap states (MIGS).[43;44;45] These states are due to hybridization between the molecule and metal (electrode generally speaking) states. Thus, we assume that the transition voltage arises when a certain amount of these states enter the energy window. Recent discussions in the literature point out how it can be possible to relate $V_{T+}$ and $V_{T-}$ to the energy position of the LUMO or HOMO levels in the junctions and to the voltage division factor γ, a parameter that describes the degree of symmetry or asymmetry of the molecular orbitals in the junction ($-0.5 \leq \gamma \leq 0.5$, $\gamma = 0$ being the case of a symmetrical coupling of the molecular orbitals between the two electrodes).[9;10;11;23] Following the analytical model of Bâldea,[11] we determine the energy level $\varepsilon_0$ of the molecular orbital involved (here LUMO) in the electrical transport (with respect to the Fermi energy of the electrodes), and γ, directly from the measured $V_{T+}$ and $V_{T-}$ according to:

$$|\varepsilon_0| = 2\frac{e|V_{T+}V_{T-}|}{\sqrt{V_{T+}^2 + 10|V_{T+}V_{T-}|/3 + V_{T-}^2}}$$
$$\gamma = \frac{sign\ \varepsilon_0}{2}\frac{V_{T+} + V_{T-}}{\sqrt{V_{T+}^2 + 10|V_{T+}V_{T-}|/3 + V_{T-}^2}} \tag{2}$$

where sign $\varepsilon_0$ is + for LUMO (- for HOMO). For the $Au_{sub}$-S-$C_nH_{2n+1}$//$Au_{tip}$ junction, we get $\varepsilon_0 = 1.12 \pm 0.10$ eV, and $\gamma \approx 0$ (see Table 1). This energy level is in very good agreement with the onset of the LUMO tail in this junction as measured by IPES[41] (starting very close to the metal Fermi energy, see Fig. 4 in ref. 36).

For the Si-$C_nH_{2n+1}$//$Au_{tip}$ junctions, the same approach for n=8 gives $\varepsilon_0 = 1.03 \pm 0.63$ eV, and $\gamma \approx 0.068$. IPES shows the LUMO edge at about 3.1 eV above the Si Fermi energy[42] but again with a clear DoS tail starting at about 1 eV[42;46] in very good agreement with this value of $\varepsilon_0$. We have to note that this comparison is meaningful because our samples were fabricated strictly following the same process as the one developed by the authors of refs 42 and 46**.**

Also note that IPES measurements[22] were carried out on long chains (16 and 18 carbon atoms) while our TVS experiments are carried out on smaller chains (n= 4 - 12). However, this is not (both experimentally and theoretically) that there is no dependence of the LUMO/HOMO gap, nor LUMO position, as function of the number of carbon atoms in the chain (at least for n larger then 4).[4;6;21;22;33] In case of π-conjugated molecules of different chemical nature measured by C-AFM ("clean junction"), Beebe et al. have shown that $V_T$ is linearly dependent on the position of the HOMO.[3;4]For all the other junctions, with one or two oxidized electrodes, we suppose that the lower values of $V_T$ ( $|V_T| < 0.6$ eV) are due to some oxide states as schematically shown in Fig. 7b. The role of these states in the electronic conduction of molecular junctions was already suggested,[47] and theoretically taken into account very recently to explain some discrepancies between experiments and TVS models.[48] Here, the role of these oxide states is confirmed by TVS measurement with C-AFM on a reference sample without SAM, i.e. a slightly oxidized silicon (~ 7.5 Å of $SiO_2$ on n-type <100> silicon, resistivity of $10^{-3}$ Ω.cm). We also measured low values, $V_{T+} = 0.09 \pm 0.05$ V and $V_{T-} = - 0.14 \pm 0.06$ V, i.e in the same range as for any other SAM junctions with oxidized electrodes. Oxide states near the Si Fermi energy have also been observed in metal-oxide-semiconductor devices, leading to low tunneling barrier heights.[49] Note that low $V_T$ values (around 0.2 V) have also been reported for other molecular junctions on Si (using acid terminated alkylthiol) and ascribed to the formation of interface states at the Si/molecule interface.[15;16]

## 5. CONCLUSION

To conclude, we have electrically characterized by TVS method a large number of various molecular junctions made with alkyl chains but with different chemical structure of the electrode/molecule interfaces. In the case of molecular junctions with “clean, unoxidized” electrode/molecule interfaces, we conclude that the TVS method allows estimating the onset of the tail of the LUMO density of states when alkylthiols and alkenes are grafted on Au and hydrogenated Si, respectively, with $|V_T|$ in the range of 0.9-1.4 V. Otherwise, in the case of “oxidized” interfaces (e.g. the same monolayer measured with Hg or eGaIn drops, or monolayers on a slightly oxidized silicon substrate), lower $|V_T|$ (0.2-0.6V) are systematically measured and related to the presence of oxide-related density of states at lower energies than the HOMO/LUMO of the molecules. As such, the TVS method is a useful technique to assess the quality of the molecule/electrode interfaces in molecular junctions.

| Junction | top contact | n | $V_{T+}$ (V) | $V_{T-}$ (V) | $\|\varepsilon_0\|$ (eV) | γ |
|---|---|---|---|---|---|---|
| $Au_{sub}$-S-$C_nH_{2n+1}$ | C-AFM | 4 | 1.40±0.24 | -1.29±0.17 | 1.16±0.25 | -0.018 |
| | | 6 | 1.34±0.10 | -1.33±0.36 | 1.16±0.33 | -0.002 |
| | | 8 | 1.21±0.13 | -1.17±0.13 | 1.03±0.16 | -0.007 |
| | | 12 | 1.33±0.04 | -1.23±0.11 | 1.11±0.10 | -0.017 |
| | | average | 1.32±0.08 | -1.26±0.07 | 1.12±0.10 | -0.010 |
| Si-$C_nH_{2n+1}$ | C-AFM | 6 | | -0.86±0.23 | -- | -- |
| | | 8 | 1.4±0.84 | -1.02±0.12 | 1.03±0.63 | -0.068 |
| | | 10 | | -1.00±0.10 | -- | -- |
| | | average | NS | -0.96±0.09 | -- | -- |
| $Au_{sub}$-S-$C_nH_{2n+1}$ | eGaIn | 4 | 0.32±0.34 | -0.42±0.14 | 0.32±0.36 | 0.059 |
| | | 6 | 0.43±0.31 | -0.59±0.04 | 0.43±0.31 | 0.068 |
| | | 8 | 0.36±0.27 | -0.48±0.04 | 0.36±0.27 | 0.062 |
| | | 12 | 0.47±0.09 | -0.52±0.25 | 0.43±0.22 | 0.022 |
| | | 14 | 0.45±0.07 | -0.52±0.07 | 0.42±0.09 | 0.031 |
| | | 16 | 0.37±0.33 | -0.58±0.12 | 0.39±0.36 | 0.096 |
| | | 18 | 0.69±0.33 | -0.52±0.25 | 0.51±0.35 | -0.061 |
| | | average | 0.44±0.12 | -0.52±0.06 | 0.41±0.12 | 0.036 |
| Si-$C_nH_{2n+1}$ | Hg drop | 10 | 0.17±0.12 | -0.43±0.05 | 0.22±0.16 | 0.192 |
| | | 14 | 0.19±0.06 | -0.19±0.06 | 0.16±0.07 | 0.000 |
| | | 18 | 0.20±0.05 | -0.25±0.07 | 0.19±0.07 | 0.048 |
| | eGaIn | 6 | 0.14±0.01 | -0.18±0.05 | 0.14±0.04 | 0.054 |
| | | 8 | 0.23±0.11 | -0.39±0.05 | 0.25±0.12 | 0.113 |
| | | 10 | 0.23±0.05 | -0.14±0.04 | 0.15±0.05 | -0.106 |
| | | 14 | 0.57±0.29 | -0.64±0.26 | 0.52±0.34 | 0.025 |
| | | 18 | 0.50±0.48 | -0.41±0.26 | 0.39±0.45 | -0.043 |
| | Al pad | 10 | 0.19±0.01 | -0.57±0.07 | 0.25±0.03 | 0.224 |
| | | 14 | 0.25±0.04 | -0.62±0.04 | 0.31±0.05 | 0.188 |
| | | 18 | 0.27±0.22 | -0.29±0.19 | 0.24±0.25 | 0.015 |
| | | average | 0.27±0.14 | -0.37±0.18 | 0.27±0.20 | 0.068 |
| Si/$SiO_2$-Si-$C_nH_{2n+1}$ | Hg drop | 8 | 0.16±0.05 | -0.16±0.05 | 0.14±0.06 | 0.000 |
| | | 12 | 0.16±0.05 | -0.16±0.05 | 0.14±0.06 | 0.000 |
| | Al pads | 8 | 0.16±0.05 | -0.12±0.01 | 0.12±0.04 | -0.062 |
| | | 12 | 0.35±0.31 | -0.19±0.02 | 0.22±0.20 | -0.130 |
| | µpore 10µm | 8 | 0.27±0.09 | -0.37±0.27 | 0.27±0.22 | 0.068 |
| | | 12 | 0.33±0.08 | -0.28±0.08 | 0.26±0.10 | -0.036 |
| | µpore 100µm | 8 | 0.17±0.02 | -0.18±0.06 | 0.15±0.05 | 0.012 |
| | | 12 | 0.17±0.02 | -0.19±0.05 | 0.16±0.04 | 0.024 |
| | eGaIn | 8 | 0.30±0.04 | -0.18±0.06 | 0.20±0.05 | -0.11 |
| | | 12 | 0.43±0.19 | -0.19±0.05 | 0.23±0.04 | -0.17 |
| | | average | 0.25±0.10 | -0.21±0.08 | 0.20±0.11 | -0.038 |
| Reference sample | eGaIn | 0 | 0.24±0.02 | -0.23±0.05 | 0.20±0.05 | -0.009 |
| | Micropore | 0 | 0.19±0.05 | -0.23±0.06 | 0.18±0.07 | 0.041 |
| | Hg drop | 0 | 0.24±0.05 | -0.13±0.01 | 0.15±0.03 | -0.130 |

**Table 1. Summary of the measured $V_{T+}$ and $V_{T-}$ for all the junctions investigated in this work (n is the number of carbon atoms in the alkyl chain). NS means "not significant". Energy level $\varepsilon_0$ and asymmetry factor $\gamma$ deduced from eqs. (2) for all the measured junctions.**

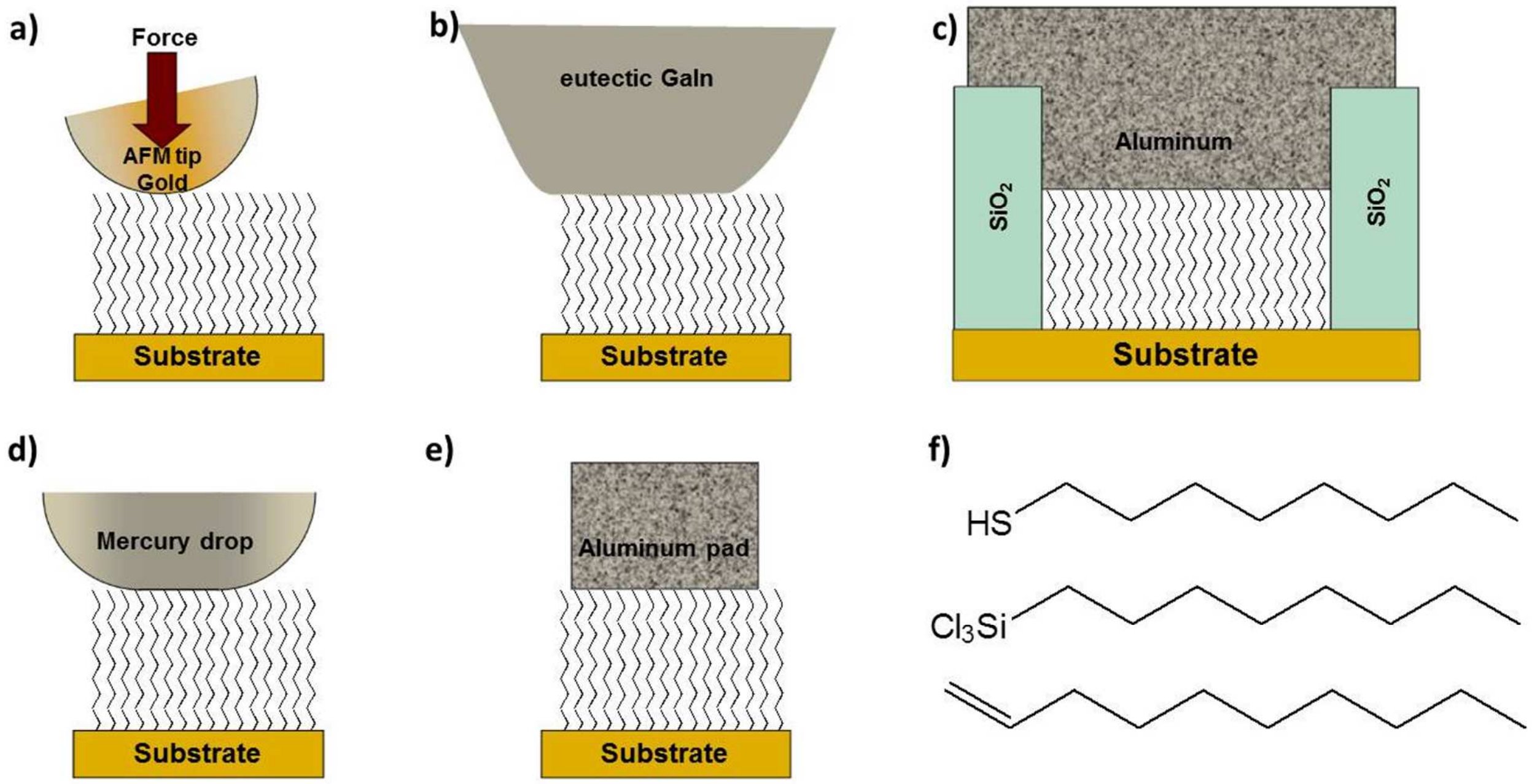

**Figure 1.** Schematic representation of the different molecular junctions. The substrate is gold, hydrogenated silicon or natively oxidized silicon. The top electrode is: a) a conducting-AFM with gold tip, b) an eutectic GaIn drop, c) an evaporated Al electrode on a SAM grafted in a micropore junction fabricated in $SiO_2$ by conventional lithography, d) a mercury drop, e) an evaporated aluminum pads through a shadow mask. f) the three families of alkyl molecules (from top to bottom): alkylthiol ($CH_3$-$(CH_2)_{n-1}$SH with n=4;6;8;12;14;16;18), alkyltrichlorosilane ($CH_3$-$(CH_2)_{n-1}SiCl_3$) with n=8; 12) and alkene ($CH_2$=CH-$(CH_2)_{n-3}$-$CH_3$ with n=6; 8; 10).

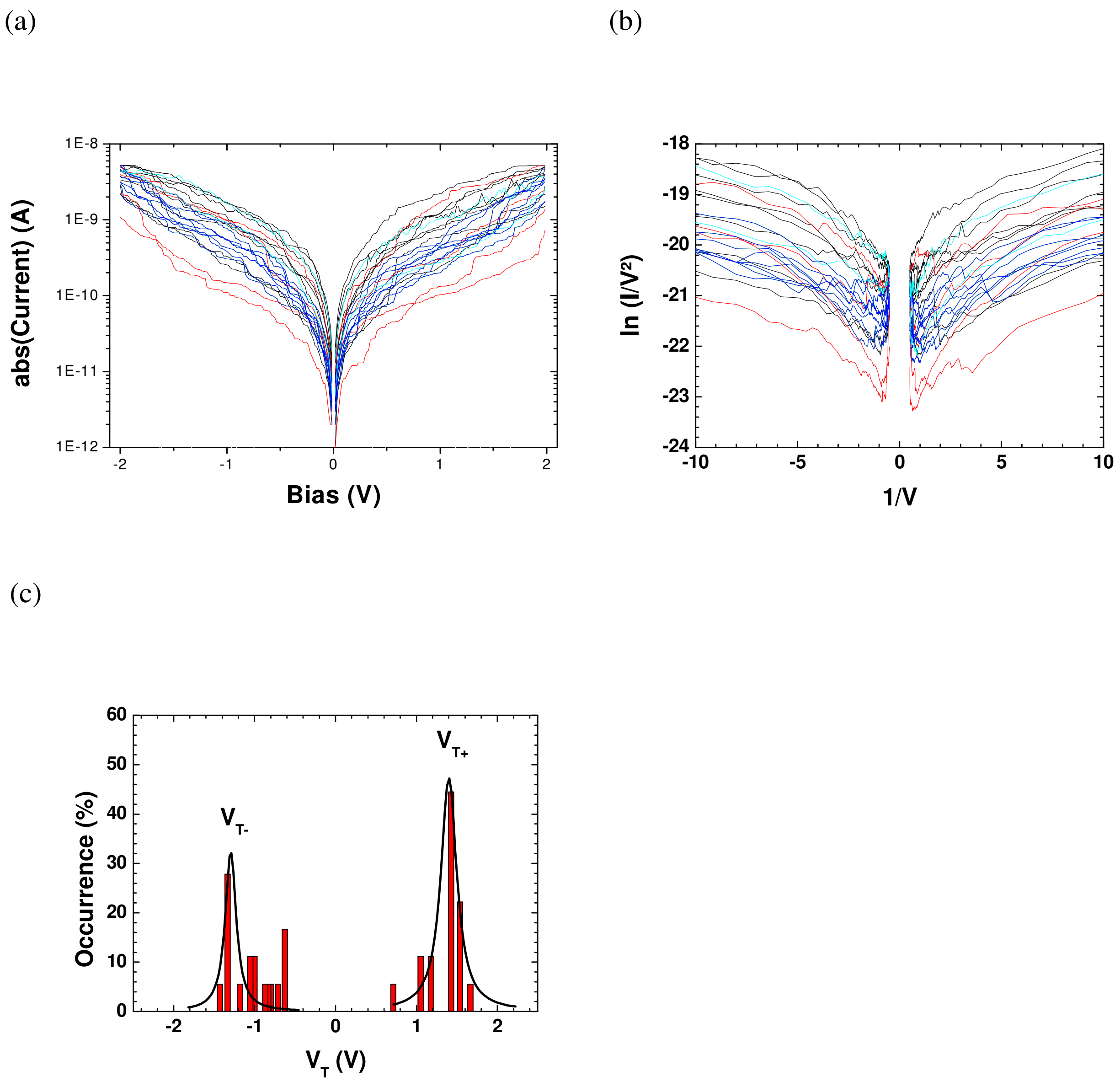

**Figure 2:** Example for the junction $Au_{sub}$-S-$C_nH_{2n+1}$//$Au_{tip}$ with n=4 of (a) the set of 19 I-V curves measured on this junction; (b) the Fowler-Nordheim plots obtained from these I-V curves and (c) $V_T$ histograms obtained from the Fowler-Nordheim plots, from which we get the $V_{T+}$, $V_{T-}$ and standard deviation values presented in table 1 (black line is the Gaussian fit).

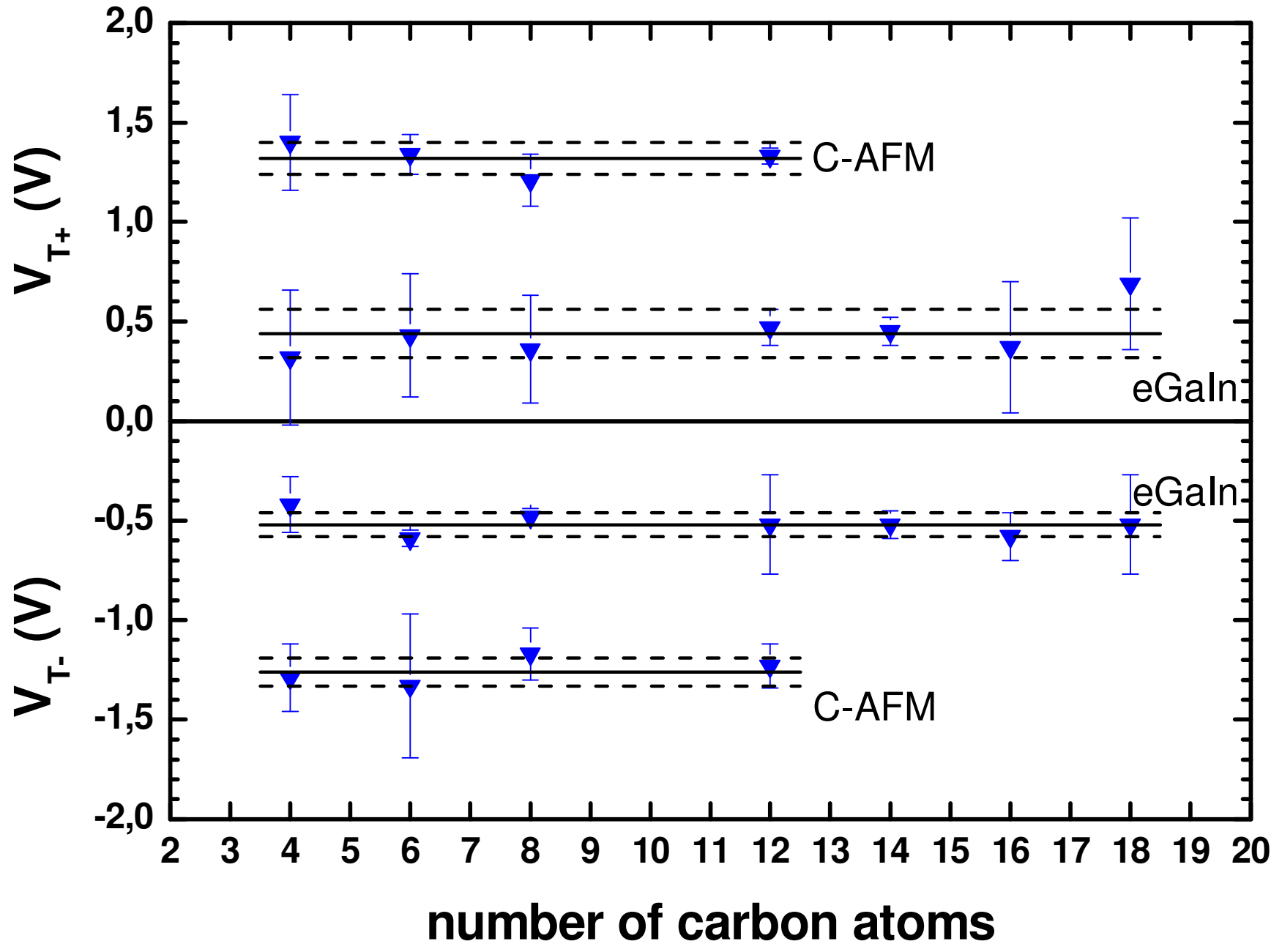

**Figure 3.** $V_{T+}$ and $V_{T_}$ values for alkylthiol SAMs determined from 5 to 20 I-V curves by C-AFM at ca. 20nN and by eGaIn contact, as function of the length of the alkyl chain. The solid lines correspond to the average value for $V_T$, and the dashed lines are the standard deviation. These average values for $V_{T+}$ and $V_{T-}$ determined for this set of junctions by C-AFM and by eGaIn drop are respectively $V_{T+ave}$ = 1.32 ± 0.08 V; $V_{T-ave}$= -1.26 ± 0.07 V and $V_{T+ave}$ = 0.44 ± 0.12 V; $V_{T-ave}$ = - 0.52 ± 0.06 V.

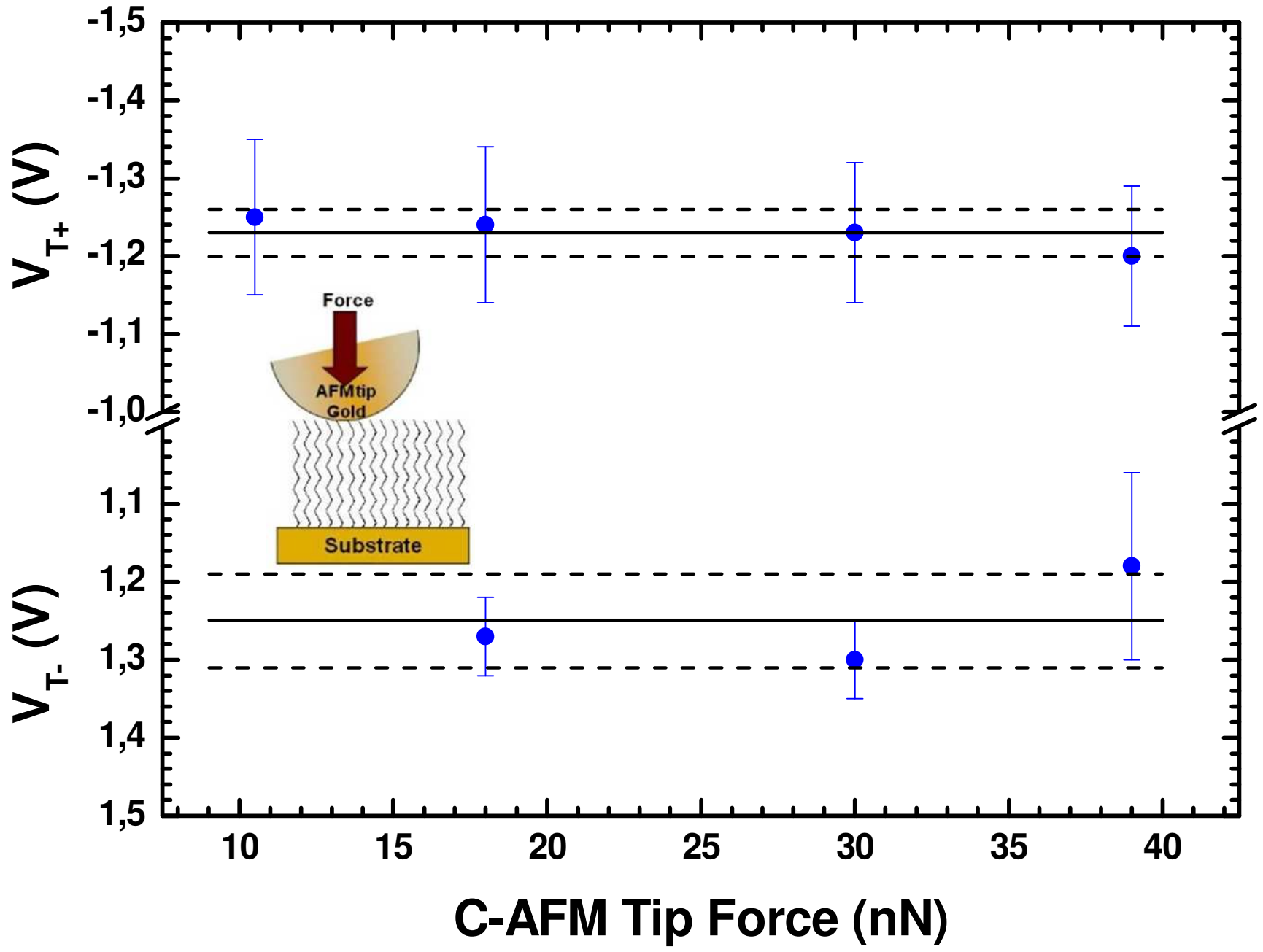

**Figure 4:** $V_{T+}$ and $V_{T-}$ determined from 5 I-V curves for the $Au_{sub}$-S-$C_nH_{2n+1}$ (n=12) junction as function of the loading force on the AFM tip. The solid lines correspond to the average value for $V_T$, and the dashed lines the standard deviation.

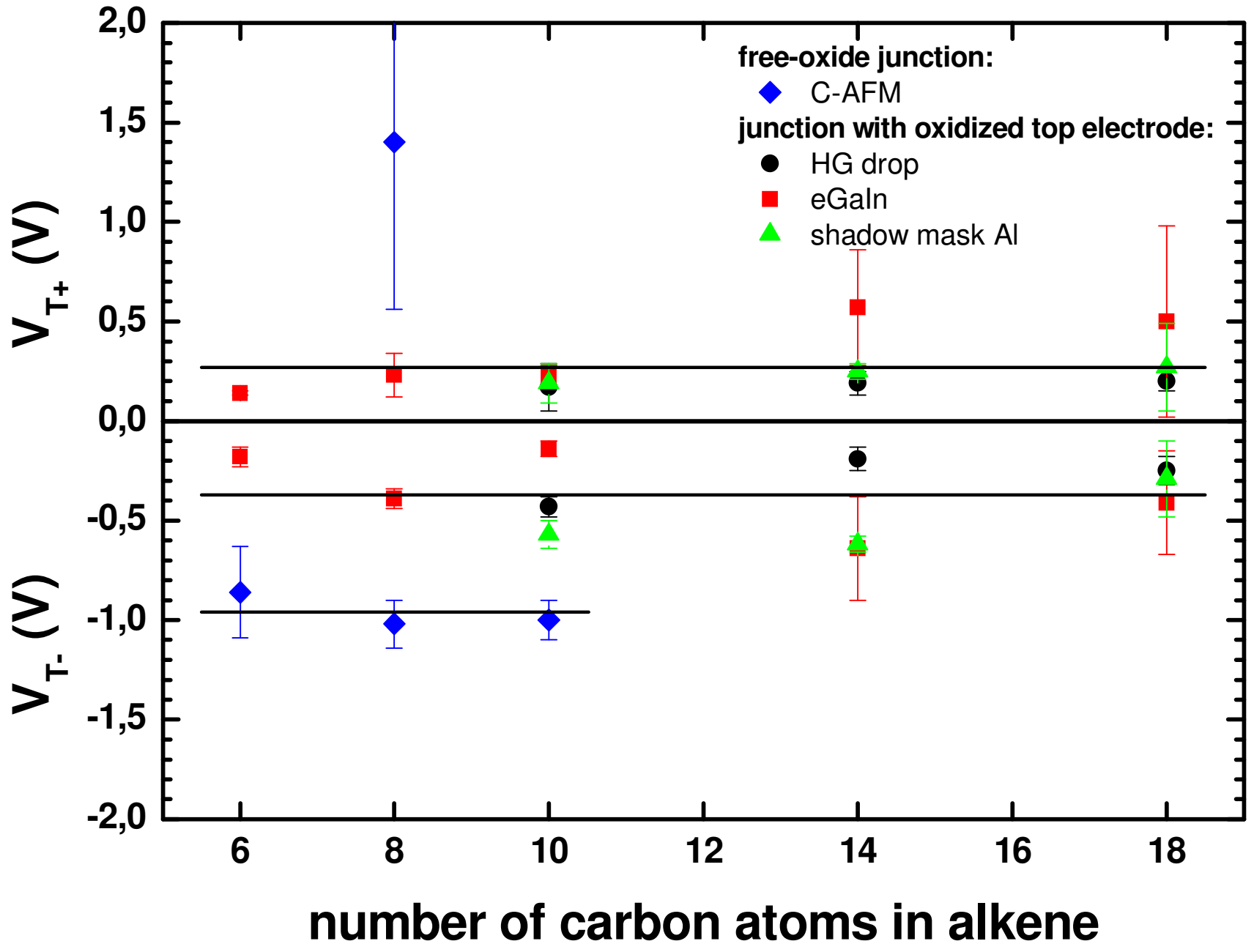

**Fig. 5:** $V_{T+}$ and $V_{T-}$ determined for the Si-$C_nH_{2n+1}$ junctions in the case of oxide-free junction (C-AFM top electrode) and for junctions with oxidized top electrode (mercury drop, eGaIn drop and aluminum patterns). Each point corresponds to an average on 5-18 I-V curves, and the error-bar to the standard deviation. For negative bias, two families of measurements are clearly distinguishable: for "clean junctions" with an averaged $V_{T\text{-ave}}$ = - 0.96 ± 0.09 V, and for "oxidized junction" $V_{T\text{-ave}}$ = - 0.37 ± 0.18 V and $V_{T\text{+ave}}$ = 0.27 ± 0.20 V as indicated by horizontal solid lines.

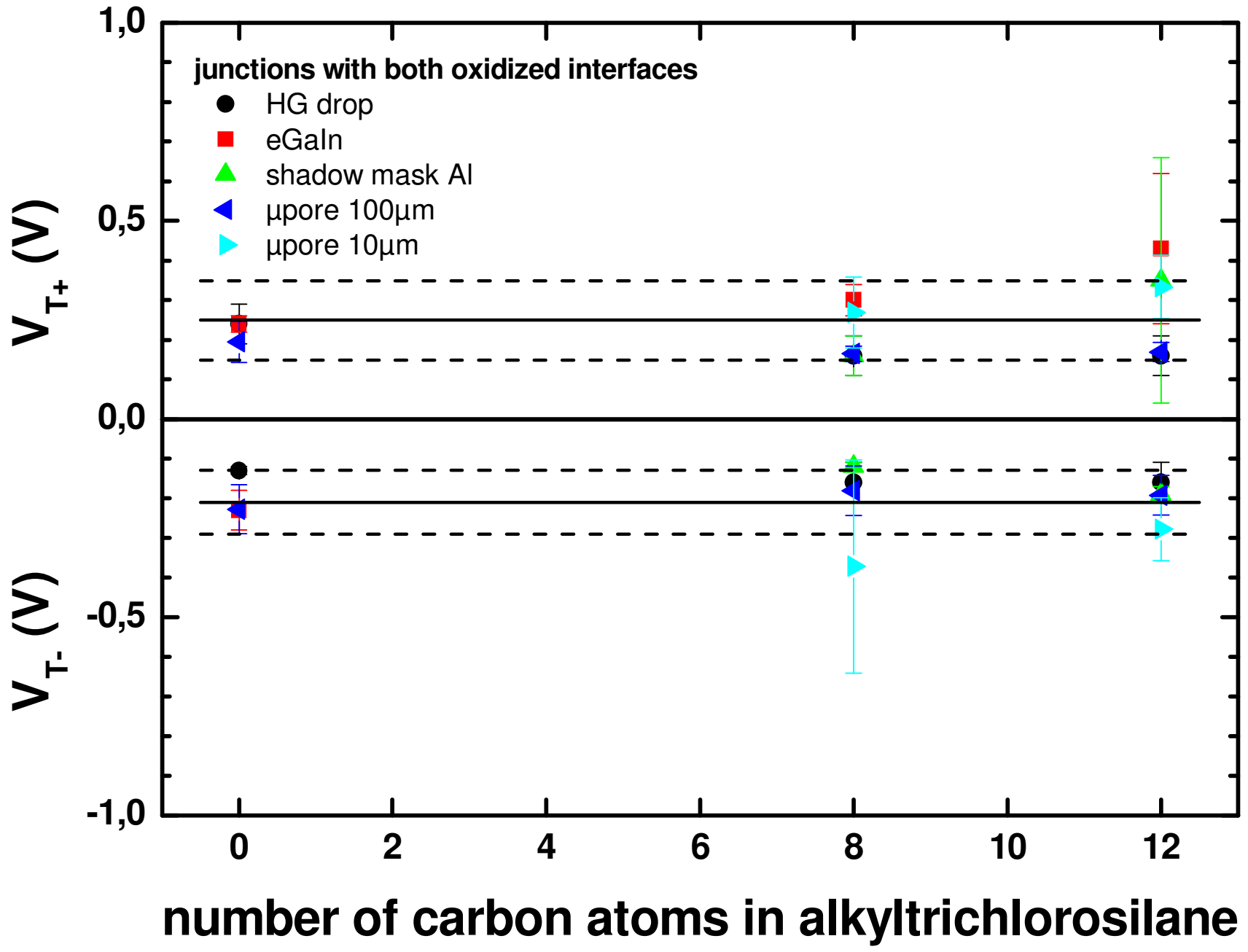

**Figure 6:** $V_{T+}$ and $V_{T-}$ determined for the Si/$SiO_2$-Si-$C_nH_{2n+1}$ junctions in the case of both oxidized interfaces: silicon with ca. 1.3nm-thick silicon dioxide and oxidized top electrode (mercury drop, eGaIn drop and aluminum patterns). Each point corresponds to an average on 5-28 I-V curves, and the error bar to the standard deviation. Whatever the nature of the metal top electrode, $V_T$ values are quite similar. Averaging over the different lengths and type of top electrodes gives: $V_{T+ave} = 0.25 \pm 0.10$ V, and $V_{T-ave} = -0.21 \pm 0.08$ V as shown by horizontal solid lines in the graph, the dashed lines being the standard deviation.

(a)

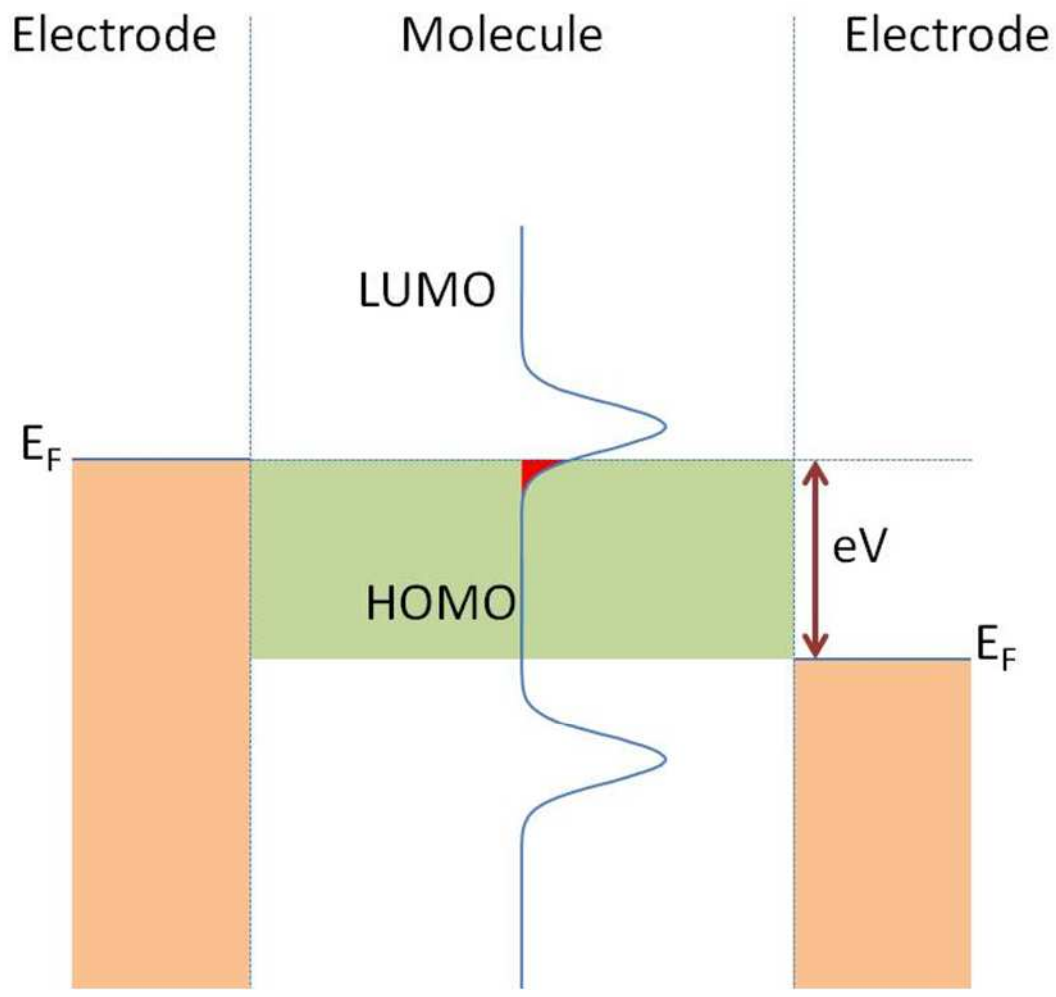

(b)

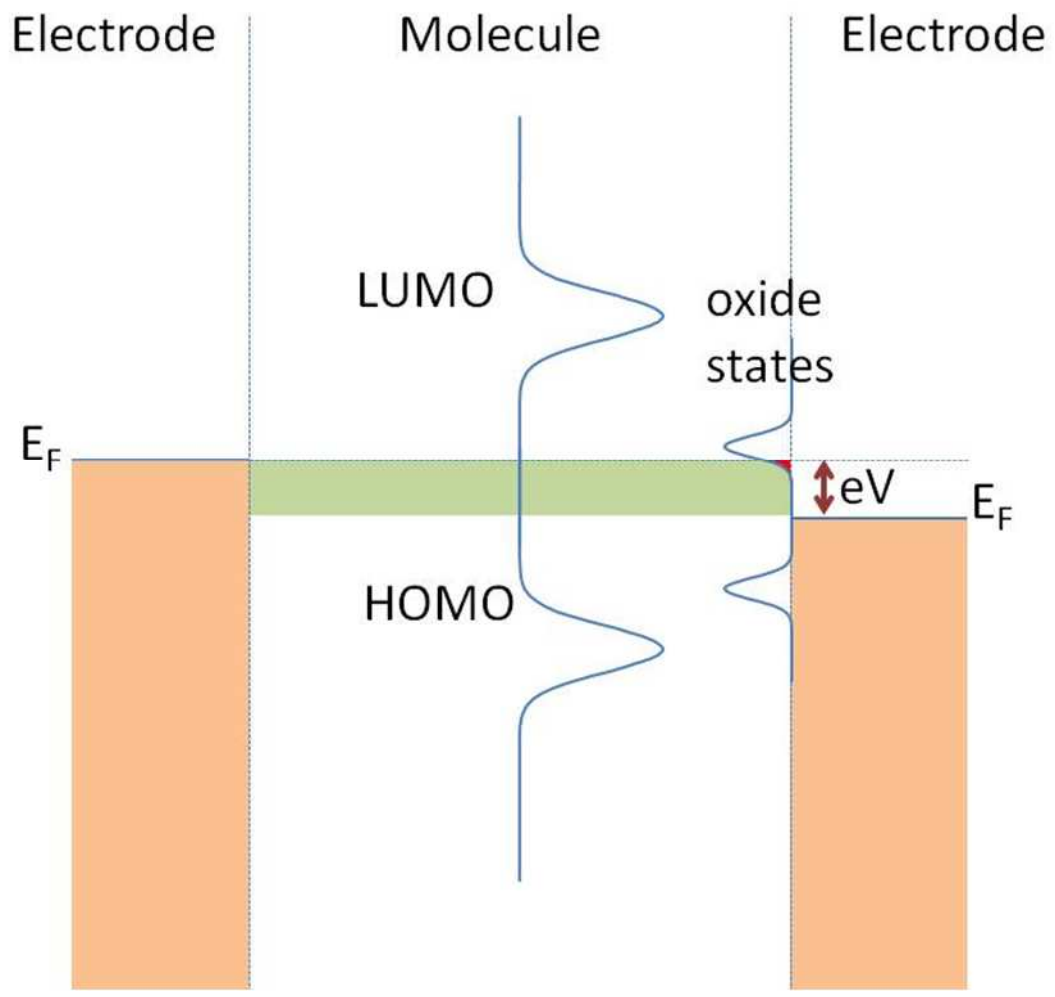

**Figure 7.** Schematic representations of the energy level diagrams at the threshold point of the TVS plot curve (*i.e.* when the applied bias is close to $V_T$) for (a) an oxide-free clean and (b) a junction with an oxidized electrode. The two electrodes are represented by the Fermi energies, the molecule by the HOMO and LUMO energies, and the oxide by energy bands close to the Fermi energy of the oxidized electrode.

ASSOCIATED CONTENT

**Supporting information**. Ellipsometry and water contact angle measurements are available free of charge via the Internet at http://pubs.acs.org.

AUTHOR INFORMATION

**Corresponding Author**

* E-mail: dominique.vuillaume@iemn.univ-lille1.fr; stephane.lenfant@iemn.univ-lille1.fr

ACKNOWLEDGMENT

G.R. thanks the CNRS and the Région Nord-Pas de Calais for the PhD grant funding (contract number 72837). We thank Nicolas Clément and Stéphane Pleutin for fruitful discussions, Dominique Deresmes for assistance in C-AFM, and the clean room staff for the assistance and guidance for micropore fabrication.

# TOC GRAPHIC

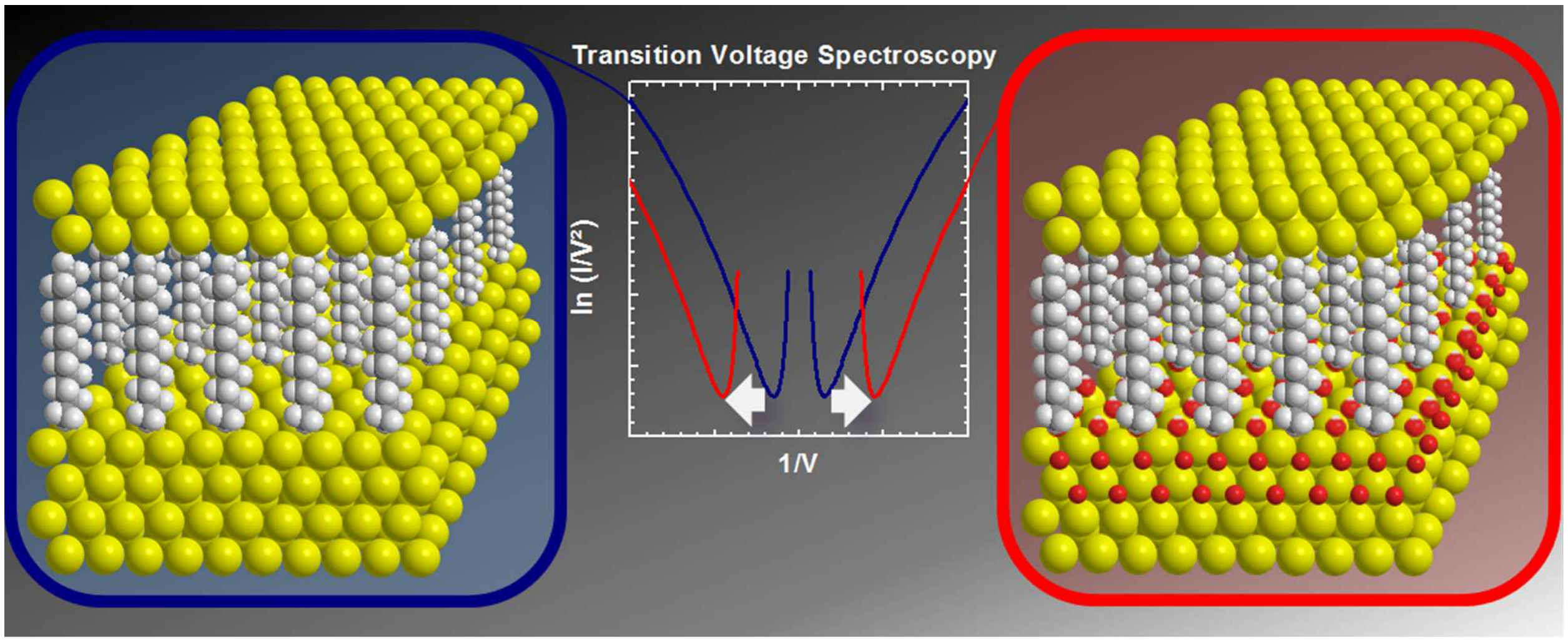

# Molecule/Electrode Interface Energetics in Molecular Junction: a "Transition Voltage Spectroscopy" Study

*Guillaume Ricœur, Stéphane Lenfant*, David Guérin, Dominique Vuillaume**

Institut d'Electronique Microélectronique et Nanotechnologie (IEMN), CNRS, University of Lille, B.P. 60069, Avenue Poincaré, F-59652, Villeneuve d'Ascq, France

* To whom correspondence should be addressed. E-mail: dominique.vuillaume@iemn.univ-lille1.fr; stephane.lenfant@iemn.univ-lille1.fr

**Supporting Information**

**Contact-Angle Measurements** We measured the water contact angle with a remote-computer controlled goniometer system (DIGIDROP by GBX, France). We deposited a drop (10-30 µL) of deionized water (18MΩ.cm$^{-1}$) on the surface, and the projected image was acquired and stored by the computer. Contact angles were extracted by contrast contour image analysis software. These angles were determined a few seconds after application of the drop. These measurements were carried out in a clean room (ISO 6) where the relative humidity (50%) and the temperature (22 °C) are controlled. The precision with these measurements are ±2°.

**Spectroscopic Ellipsometry** We recorded spectroscopic ellipsometry data in the visible range using an UVISEL (Jobin Yvon Horiba) spectroscopic ellipsometer equipped with a DeltaPsi 2 data analysis software. The system acquired a spectrum ranging from 2 to 4.5 eV (corresponding to 300-750 nm) with intervals of 0.1 eV (or 15 nm). Data were taken at an angle of incidence of 70°, and the compensator was set at 45.0°. We fitted the data by a regression analysis to a film-on-substrate model as described by their thickness and their complex refractive indexes. First, we recorded a background before monolayer deposition. Second, after the monolayer deposition, we used a two-layer model (substrate/SAM) to fit the measured data and to determine the SAM thickness. We used the previously measured optical properties of the substrate (background), and we fixed the refractive index of the organic monolayer at 1.50. The usual values in the literature for the refractive index of organic monolayers are in the range 1.45 - 1.50 [1;2]. We can notice that a change from 1.50 to 1.55 would result in less than 1 Å error for a thickness less than 30 Å. We estimated the accuracy of the SAM thickness measurements at ± 1 Å.

Water contact angle and ellipsometric thicknesses measurements are presented below for the different SAMs:

| **Molecule / n carbon** | | **4** | **6** | **8** | **10** | **12** | **14** | **16** | **18** | **Ref.** |
|---|---|---|---|---|---|---|---|---|---|---|
| Alkyl-Thiol | e (Å) | 4.2 | 6.1 | 8.4 | | 11.6 | 15.2 | 15.7 | 23.7 | |
| | Θ (°) | 96° | 102° | 107° | | 109° | 110° | 79° | 108° | 99-108° [3] |
| Alkyl-Trichlorosilane | e (Å) | | | 12.6 | | 15.3 | | | 27 | |
| | Θ (°) | | | 108° | | 100° | | | 112° | 100-110° [4] |
| Alkene | e (Å) | | 9.1 | 10.4 | 11 | | 16.2 | | 18.2 | |
| | Θ (°) | | 97 | 100 | 107 | | 108 | | 109 | 111° [5] |

**Table S1**: Thicknesses measured by ellipsometry and water contact angles

For comparison, water contact angle measured on same self-assembled monolayers are presented with the associated reference. Thickness values are also presented in the graph below as a function of the number of carbon atom.

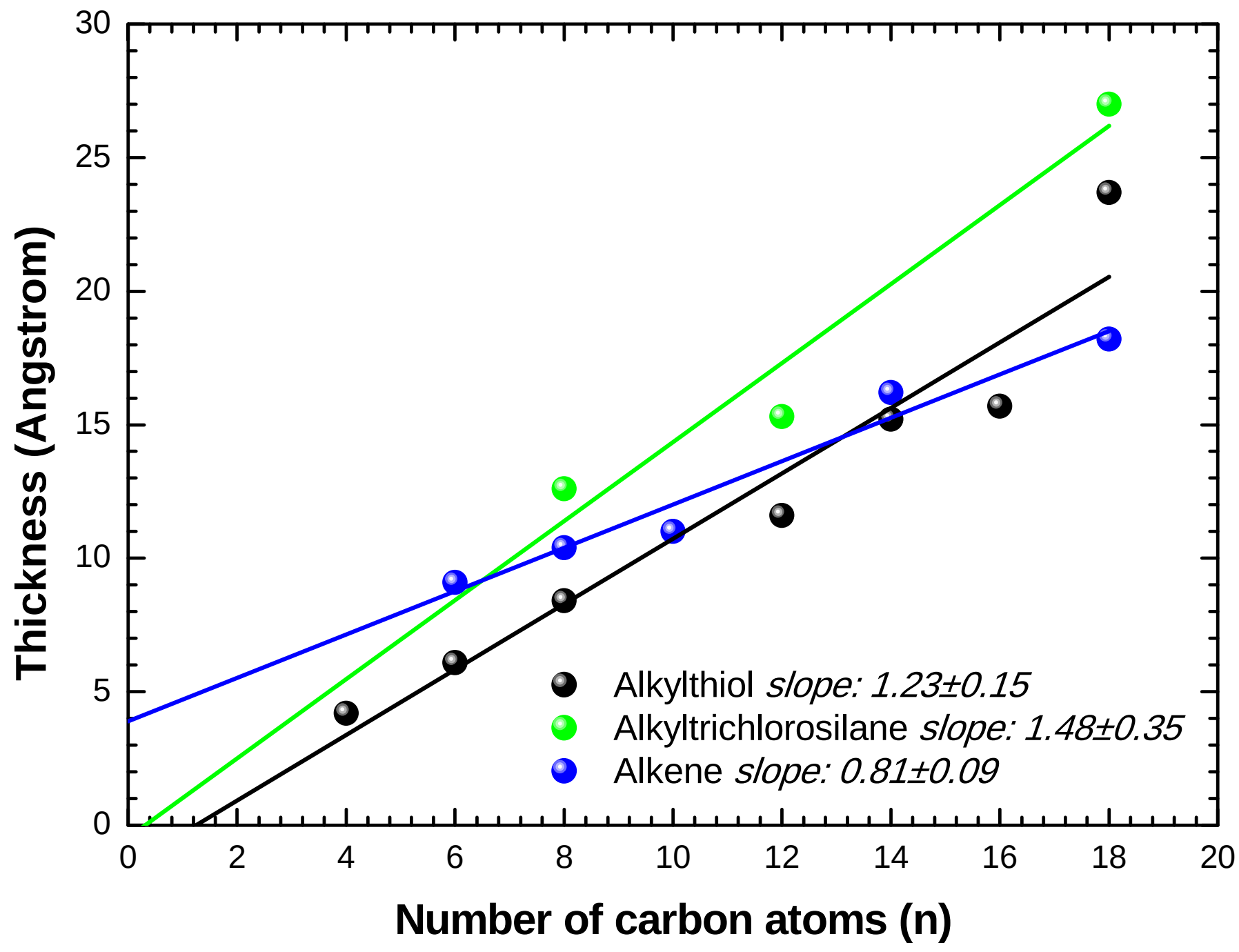

For alkyltrichlorosilane monolayers, the dependence with n the number of carbon atoms (the slope from the graph) is 1.48 ± 0.35 Å/methylene, a value closed to the one obtained by *Wasserman et al.* [4] (1.26 Å/methylene). For alkylthiol monolayers on gold surfaces, *Porter et al.* [6] measured a slope of 1.5 Å/methylene, close to the value measured here (1.23 ± 0.15 Å/methylene). In the case of alkene monolayer grafted on hydrogenated silicon, the slope measured here (0.81 ± 0.09 Å/methylene) is also close to the value obtained by Seitz and coworkers (1.0 Å/methylene) [5].